\documentclass{icrc2009}

\usepackage{graphicx}   
\usepackage[caption=false]{caption}    
\usepackage[font=footnotesize]{subfig} 
\usepackage{fixltx2e}
\usepackage{url}

\newcommand{\shorttitle}[1]%
{\markboth{Proceedings of the 31\MakeLowercase{$^{st}$} ICRC, {\L}\'{o}d\'{z} 2009}{#1} }

\begin{document}
\title{Detection of the crab pulsar with MAGIC}

\author{  \IEEEauthorblockN{M. Lopez\IEEEauthorrefmark{1},
  N. Otte\IEEEauthorrefmark{2}\IEEEauthorrefmark{3},
  M. Rissi\IEEEauthorrefmark{4},
  T. Schweizer\IEEEauthorrefmark{2},
  M. Shayduk\IEEEauthorrefmark{2},
  and S. Klepser\IEEEauthorrefmark{5} \\
  for the MAGIC collaboration}
                            \\
\IEEEauthorblockA{\IEEEauthorrefmark{1} Universit\`a di Padova and INFN,  I-35131 Padova, Italy}
\IEEEauthorblockA{\IEEEauthorrefmark{2} Max-Planck-Institut f\"ur Physik, D-80805 M\"unchen, Germany}
\IEEEauthorblockA{\IEEEauthorrefmark{3} now at: University of California, Santa Cruz, CA 95064, USA}
\IEEEauthorblockA{\IEEEauthorrefmark{4} ETH Zurich, CH-8093 Switzerland}
\IEEEauthorblockA{\IEEEauthorrefmark{5} IFAE, Edifici Cn., Campus UAB, E-08193 Bellaterra, Spain}
}

\shorttitle{Detection of the crab pulsar with MAGIC}
\maketitle

%
%

\begin{abstract}
The MAGIC telescope has detected for the first time pulsed gamma-rays from the
 Crab pulsar in the VHE domain \cite{CrabScience}. 
 The observations were performed with a newly developed trigger system that
 allows us to lower the energy threshold of the telescope from 55 GeV to 25
 GeV.
We present a
comparison of light curves measured by our experiment with the one measured by space detectors. A strong energy dependent decrease of the first peak with respect to the second peak P1/P2 could be observed.
Finally, fitting our measured data and previous measurements from EGRET we determine a turnover of the energy spectrum at 17.7 +- 2.8 (stat.) +- 5.0
(syst.) GeV, assuming an exponential cutoff.  This rules out the scenario in which the gamma rays are produced in
vicinity of the polar caps of the neutron star.

 \end{abstract}

\begin{IEEEkeywords}
Pulsars, Crab, IACTs
\end{IEEEkeywords}

%
%
\section{Introduction}

\begin{figure*}[th]
  \centering
  \includegraphics[width=.65\linewidth]{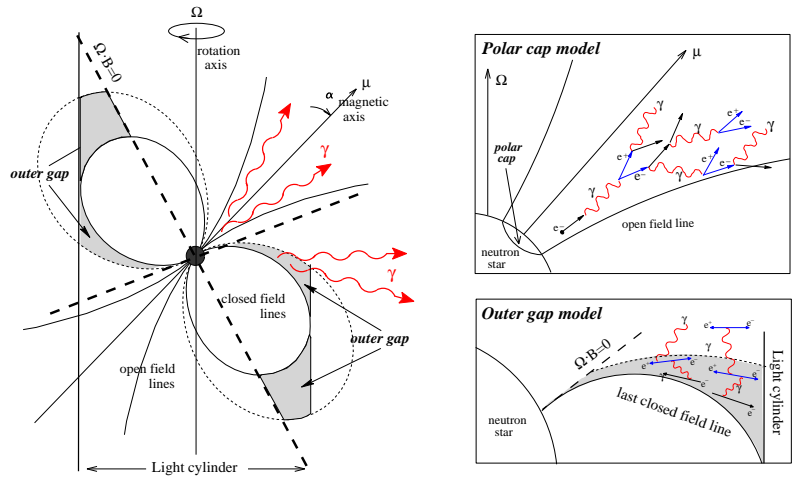}
 \caption{A sketch of a pulsar's magnetosphere (left) and illustration of the most
  populars $\gamma$-ray emission mechanisms (right).}
\label{fig_models}
 \end{figure*}

The mechanism of the pulsed electromagnetic emission in the Crab pulsar is still an open
fundamental question. 
Observations with the EGRET instrument  on-board {\it Compton Gamma-Ray
  Observatory} \cite{EGRET}
 led to the detection of the Crab pulsar up
to energies of $\sim 10$ GeV, in addition to other six $\gamma$-ray pulsars and a few more
likely candidates \cite{Thompson} (recently confirmed by the Fermi Gamma-ray
Space Telescope \cite{FermiCatalog}). In their turn, all the groups operating
Cherenkov telescopes have been trying during the last 30 years to detect the
Crab pulsar without success, being only the steady emission coming from its
nebula visible at TeV energies. This suggested that the Crab pulsed spectrum should
terminate at energies of tens of GeV. Although the existence of a sharp cutoff
in the spectrum of pulsars is a common prediction of the different theoretical models, the energy at which this
cutoff happens and its spectral features change from model to model. In the
polar cap model (see e.g. \cite{PolarCap}), electrons are accelerated
above the polar cap radiating $\gamma$-rays via synchro-curvature radiation.
Since these $\gamma$-rays are created in superstrong magnetic
fields, magnetic pair production is unavoidable, and hence, only those secondary photons which
survive pair creation (a few GeV for typical pulsars) escape to infinity as an
observed pulsed emission (see Fig. \ref{fig_models}). A natural consequence of the polar cap process is a
superexponential cutoff of the spectrum above a characteristic energy $E_0$. In
the outer gap model \cite{OuterGap} $\gamma$-ray production is expected to occur near the
 light cylinder of the pulsar, far away from the stellar surface. In this case the cutoff is determined by photon-photon
pair production, which has a weaker energy dependence compared to magnetic pair
production, and therefore a higher energy cutoff may be observable.

Thanks to its low energy threshold, MAGIC is the first ground-based
 $\gamma$-ray telescope able to overcome the sharp cutoffs
 expected near 10 GeV and detect pulsed $\gamma$-rays. This allows to
 measure the spectral shape of the pulsed emission in the relevant energy
 range, and therefore to discriminate between different emission models.

%
%
\section{The MAGIC Telescope}

The 17 m diameter MAGIC (Major Atmospheric Gamma Imaging Cherenkov) telescope
is a state of the art instrument for exploring the very high energy $\gamma$-ray
Universe. It is located  on
 the Roque de los Muchachos Observatory, at La Palma island (Spain).  
 MAGIC was built and is operated by a large international collaboration, including
 about 150 researchers.  
A $\gamma$-ray source emitting at a flux level of 1.6\% of the Crab Nebula
can be detected at a 5 $\sigma$ significance level in 50 hours of observations. 
The relative energy
resolution above 100 GeV is better than 30\% and the angular resolution is $\sim$ 0.1$^{\circ}$.
 The construction of a second telescope is now in its final stage and MAGIC
 will start stereoscopic observations in the coming months.

\begin{figure}[ht]
     \centering
 \includegraphics[width=0.35\textwidth]{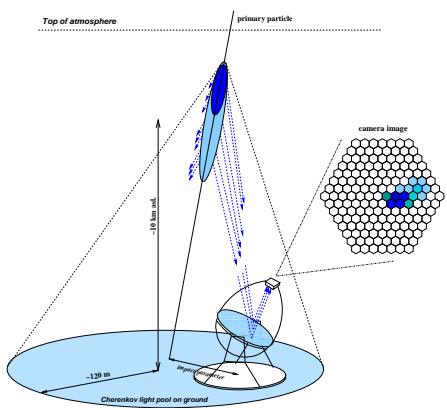}
\caption{The telescope 17 m MAGIC telescope detects
  $\gamma$-rays through short light flashes that are produced when
  $\gamma$-rays cross the atmosphere. An image of the shower
  is obtained in the telescope camera.}
\label{fig_imaging}
\end{figure}

MAGIC works by detecting the faint flashes of Cherenkov light produced when 
 $\gamma$-rays (or cosmic-rays) plunge into the earth atmosphere and initiate
 showers of secondary particles. The Cherenkov light emitted by the charged
 secondary particles is reflected by the telescope mirror and an image of the
 shower is obtained in the telescope camera (see Fig. \ref{fig_imaging}). 
An offline analysis of the shower images allows the rejection of the hadronic
cosmic ray background, the measurement of the incoming direction of the
$\gamma$-rays, and the estimation of their energy.

The MAGIC telescope was built with
the aim of achieving the lowest possible energy threshold, and since 2004
it operates with the lowest threshold worldwide, namely $\sim 50$ GeV. However, even
this low threshold turned out to be  too high  to get a clear  
signal from the Crab pulsar \cite{magicCrab}. This lead the MAGIC collaboration to build an
innovative trigger concept aimed at lowering the
threshold by a factor of 2. This new trigger is based on the analogue
summation of the signals coming from clusters of 18 pixels, instead of
discriminating single PMT signals (as it is done in the MAGIC standard
trigger). At low energies, this approach provides a better
discrimination of the faint flashes of Cherenkov photons from
the night sky background \cite{sumtrig}.

%
%

\section{Observations and data analysis}

The observations of the Crab pulsar with the new trigger system were performed
between October 2007 and February 2008. 
Together with
each event image we recorded the absolute arrival time of the corresponding
cosmic-ray with a precision of better than 1 $\mu$s from a GPS receiver, and we
recorded simultaneously also the optical signal of the Crab pulsar with a special PMT
located at the camera center \cite{cpix}. 
After rejection of data taken under unfavorable weather conditions, 22.3 hours
of  observation remained for the analysis.  We processed the data with three
independent  analysis chains, which all  gave consistent results.
In the analysis, each shower image is cleaned to remove the influence of the
night sky background, and parameterized to describe its main features. One image parameter is
the brightness of the image (SIZE) in photoelectrons, which is a good estimator
of the energy of the primary particle. Other  parameters are the orientation of
the image with respect to the source position in the camera (angle ALPHA), and
several additional parameters, which describe the shape of the image.  We apply
soft hadron rejection cuts, consisting basically in a cut in SIZE to
select only low energy showers, and a SIZE dependent cut in ALPHA optimized on
simulated Monte Carlo $\gamma$-ray events. 
For the search of pulsed emission, the arrival time of each event was
transformed to the barycenter of the solar system, 
and the corresponding  rotational phase of the Crab pulsar where calculated
using contemporaneous ephemeris provided by the Jodrell Bank Radio Telescope \cite{jodrell}.

%
\section{Results and discussion}

\begin{figure}[ht]
\centering
\includegraphics[width=0.45\textwidth]{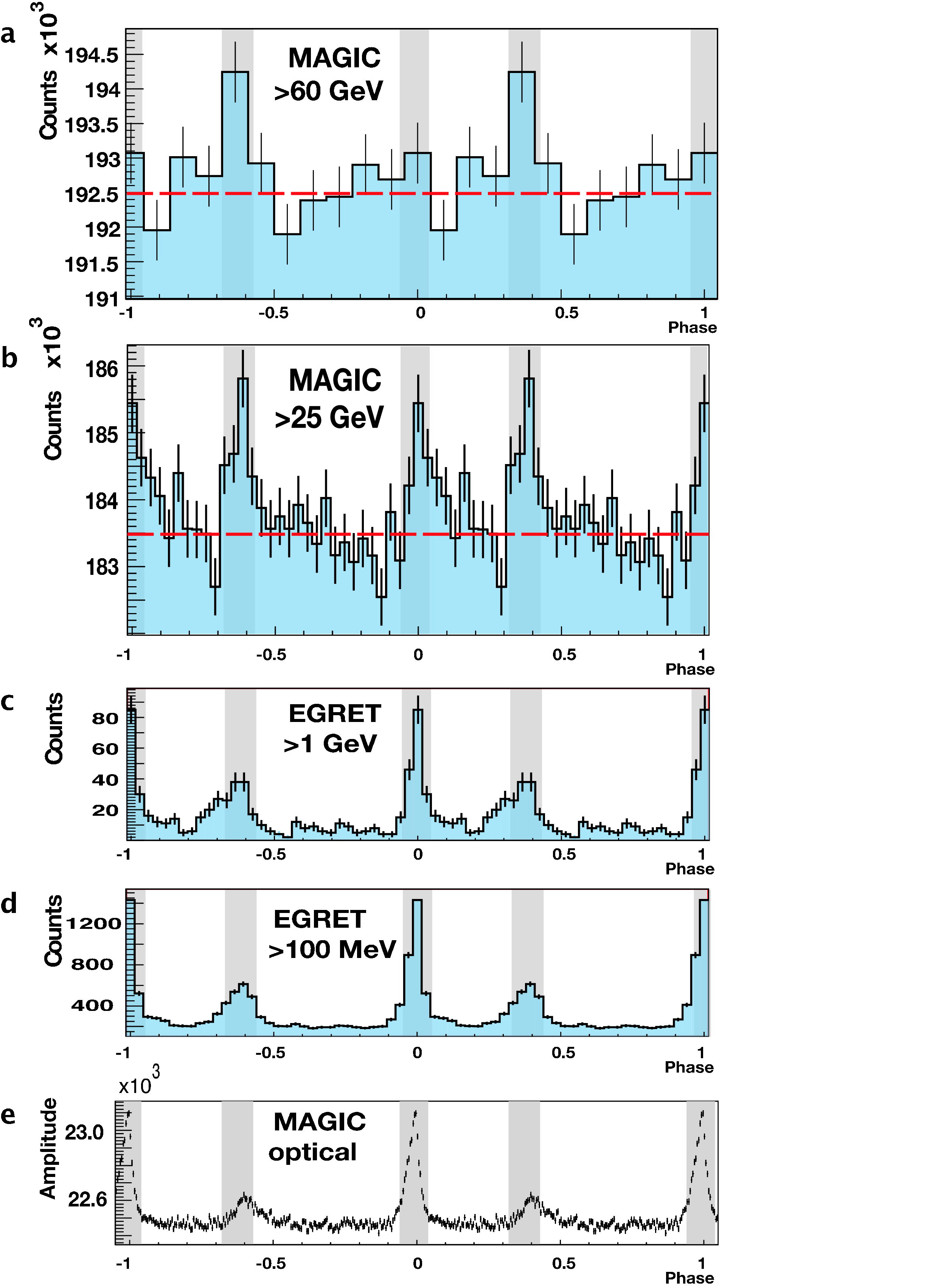}
\caption{Crab pulsed emission in different energy bands. The shaded areas show
  the signal regions for P1 and P2, as defined in \protect\cite{Fierro98}. The optical
  emission measured by MAGIC with its central pixel was recorded simultaneously
  with the $\gamma$-rays. P1 and P2 are in phase for all shown energies and the ratio of P2/P1 increases with energy from (D) to (A).}
\label{fig_lc}
\end{figure}

In figure \ref{fig_lc} we compare our pulse phase profiles in $\gamma$-rays
above  25 GeV and in the optical waveband with the measurements from the EGRET
instrument  above 100 MeV. In all profiles a pronounced signal is visible at the
position of the main pulse (at phase 0) and at the position of the inter pulse.
The significance of the pulsed signal in the $\gamma$-ray data was evaluated by
three different methods. The first method is a single hypothesis test and
assumes that $\gamma$-ray emission is expected in two phase intervals around
the main pulse and inter pulse, respectively. For the selection of the two signal
intervals we adopt the definition of the main pulse  (phase -0.06 to 0.04) and
inter pulse (phase 0.32 to 0.43) given by  \cite{Fierro98}. The background is
estimated from the remaining events outside of the intervals. In this way we
obtain a significance  of 6.4 $\sigma$. The other tho methods are
uniformity tests: the H-Test  \cite{HTest} (a periodicity test that is commonly
used for periodicity searches) and the well known Pearson's $\chi^2$ that tests
the null hypothesis that the pulse profile  follows a uniform distribution,
both given a similar significance.

\begin{figure}[ht]
 \centering
 \includegraphics[width=0.5\textwidth]{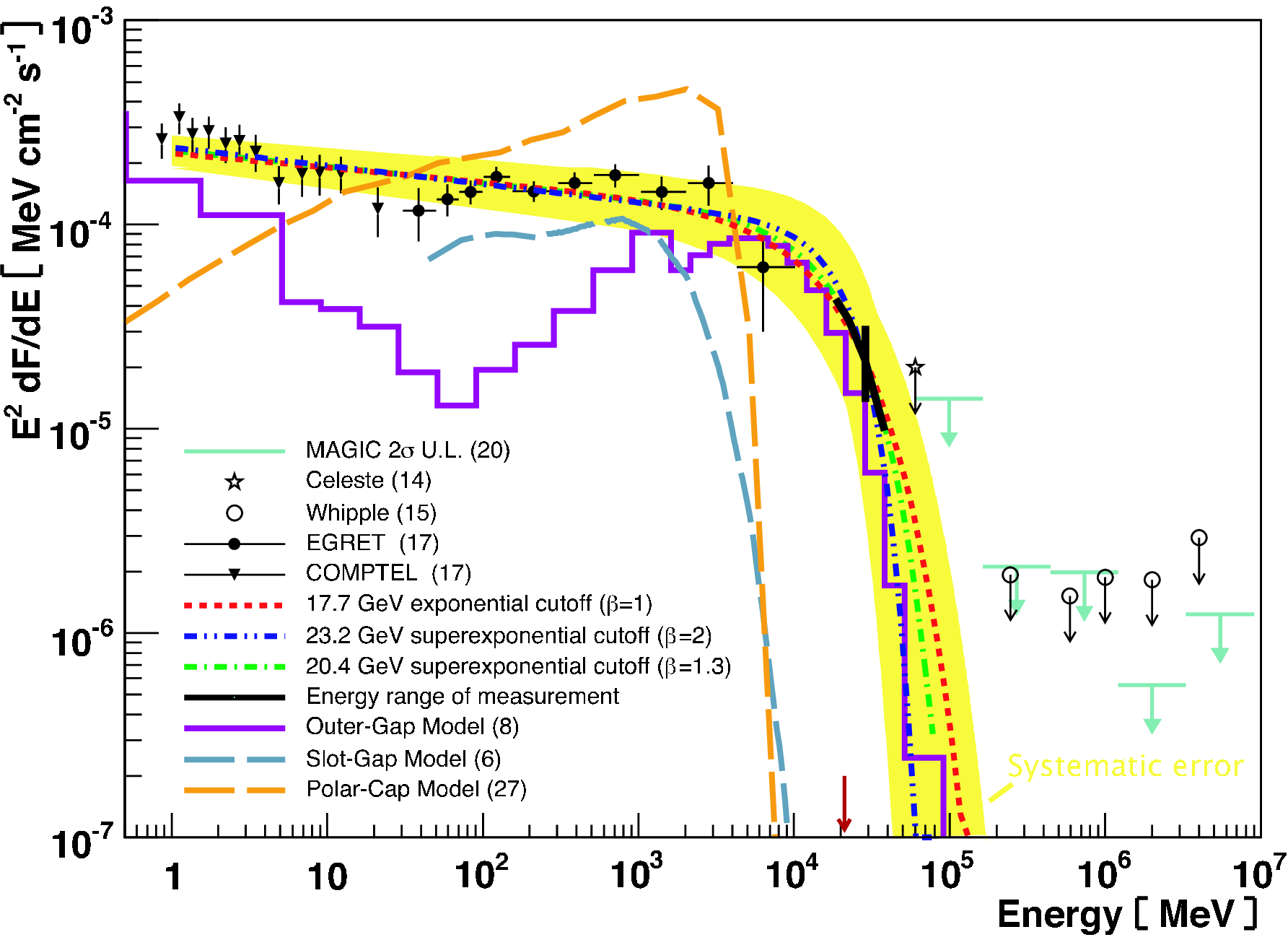}
 \caption{Crab pulsar spectrum. The solid circles and triangles on the
 left represent flux measurements from EGRET, while the arrows on the right
 denote upper limits from various previous experiments. We performed a joint
 fit of a generalized function {F(E) =$AE^{-a} exp[-(E/E0)b]$} to the MAGIC and
 EGRET data. The figure shows all three fitted functions for b = 1 (red line),
 b = 2 (blue line), and the best-fit, b = 1.2 (green line). The black line
 indicates the energy range, the flux, and the statistical error of our
 measurement. The yellow band illustrates the joint systematic error of all
 three solutions.}
\label{spectrum}
 \end{figure}

To evaluate the cutoff
energy we extrapolate the energy spectrum measured by EGRET (between 100 MeV
and 1 GeV) \cite{Fierro98} to higher energies, assuming two different
cutoff  shapes. If we assume an exponential cutoff (Flux$\times
\exp(-E/E_{0})$), the measured signal is compatible with a cutoff energy
$E_{0}$ of $17.7\pm2.8_{\rm{stat}}\pm5.0_{\rm{sys}}$ GeV.  In case the cutoff is superexponential (Flux$\times
\exp(-E/E_{0})^2$)  we determine  a cutoff energy  of
$23.2\pm2.9_{\rm{stat}}\pm6.6_{\rm{sys}}$ GeV. Figure \ref{spectrum} shows the
Crab pulsar spectrum with the cutoffs obtained in this work, compared to different theoretical predictions.
The values obtained for the
cutoff energy are higher than expected, which allow us to draw important
conclusions about the mechanism of $\gamma$-ray emission in the Crab pulsar. Using equation 1 of
\cite{Baring} that relates the location of the emission region, $r$, with the cutoff
energy, one obtains for the polar cap scenario $r/R_0>6.2\pm0.2_{\rm{stat}}\pm0.4_{\rm{sys}}$ (where $R_0$ is the
neutron star radius). This contradicts the basic picture of polar cap scenarios in
which $\gamma$-rays are emitted very close to the pulsar surface.

%
%
\section{Summary}

We succeeded in the detection of pulsed $\gamma$-rays from the Crab pulsar with
the MAGIC Cherenkov telescope above 25 GeV. This ends a 30 year-long effort of
ground based $\gamma$-ray instruments  to detect a pulsar at VHE
$\gamma$-rays. The detection was made possible by the upgrading of the trigger system,
which reduced substantially the trigger threshold from about 50 GeV to about 25
GeV. The significance of the pulsed signal is 6.4 $\sigma$. We   find that the main pulse and inter pulse in the pulse profile have about equal peak amplitudes in our energy range.
 We determine the
cutoff in the energy spectrum at  $17.7\pm2.8_{\rm{stat}}\pm5.0_{\rm{sys}}$ GeV assuming that the cutoff is exponential
in shape. The cutoff energy shifts to  $23.2\pm2.9_{\rm{stat}}\pm6.6_{\rm{sys}}$  GeV if the cutoff is
superexponential. The high value of the cutoff, and a marginally better fit
with a simple exponential point to an acceleration region located at high
altitude in the magnetosphere.\\

%
%
\section*{Acknowledgments.} The support of the German BMBF and MPG, the Italian
INFN, 
the Spanish CICYT, ETH research grant TH-34/04-3, and the Polish MNiI grant 1P03D01028 is gratefully acknowledged. 
We thank also the IAC for the excellent working conditions at the ORM in La Palma.

%
%

\end{document}